%% file: Ember.tex
\documentclass[electronic, preprint]{vgtc}                          

\ifpdf
  \pdfoutput=1\relax                   
  \usepackage{graphicx}                
  \DeclareGraphicsExtensions{.pdf,.png,.jpg,.jpeg} 
\else
  \usepackage{graphicx}                
  \DeclareGraphicsExtensions{.eps}     
\fi%

\graphicspath{{figures/}{pictures/}{images/}{./}} 

\usepackage{microtype}                 
\PassOptionsToPackage{warn}{textcomp}  
\usepackage{textcomp}                  
\usepackage{mathptmx}                  
\usepackage{times}                     
\usepackage{cite}                      
\usepackage{tabu}                      
\usepackage{booktabs}                  

 \usepackage{balance}
\onlineid{0}

\vgtccategory{Research}

\vgtcinsertpkg

\preprinttext{This article was accepted to the 2022 \href{https://vis4good.github.io/}{Visualization for Social Good workshop} held as part of \href{http://ieeevis.org/year/2022/welcome}{IEEE VIS 2022}}

\title{Envisioning Situated Visualizations of Environmental Footprints\\in an Urban Environment}

\author{Yvonne Jansen\thanks{Contact author:  yvonne.jansen@cnrs.fr\\ \hspace*{4.5mm} DOI: \href{https://doi.org/10.5281/zenodo.7053934}{10.5281/zenodo.7053934}}, %
      Federica Bucchieri\textsuperscript{$\dag$}, %
      Pierre Dragicevic*, %
      Martin Hachet*, %
      Morgane Koval*,\\
      Léana Petiot*, 
      Arnaud Prouzeau*, 
      Dieter Schmalstieg\textsuperscript{\textopenbullet}, 
      Lijie Yao\textsuperscript{$\dag$}, 
      and Petra Isenberg\textsuperscript{$\dag$}\\[1ex]%
        \small *Univ.\@ Bordeaux, CNRS, Inria, LaBRI, France\\\small \textsuperscript{\dag}Université Paris-Saclay, CNRS, Inria, LISN, France\\\small \textsuperscript{\textopenbullet}Graz University of Technology, Austria}

\teaser{
  \centering
  \includegraphics[width=\linewidth]{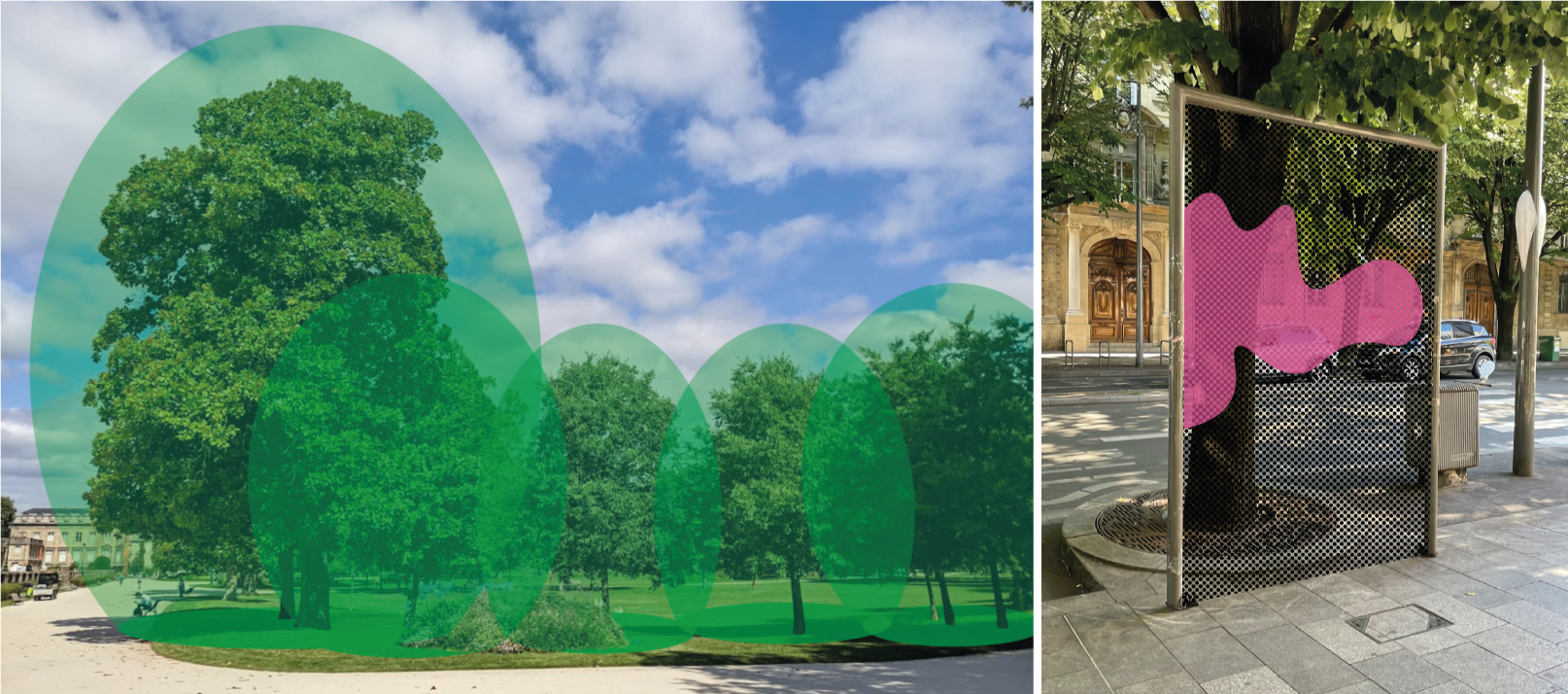}
  \caption{Examples of situated visualizations envisioned during the workshop. On the left, each green bubble represents the positive effect the trees have on the atmosphere and on oxygen production. On the right, the proportion of carbon dioxide in the air is displayed and updated every time a vehicle passes behind.}
  \label{fig:teaser}
}

\abstract{ %
We present the results of a brainstorming exercise focused on how situated visualizations could be used to better understand the state of the environment and our personal behavioral impact on it. Specifically, we conducted a day long workshop in the French city of Bordeaux where we envisioned situated visualizations of urban environmental footprints. We explored the city and took photos and notes about possible situated visualizations of environmental footprints that could be embedded near places, people, or objects of interest. We found that our designs targeted four purposes and used four different methods that could be further explored to test situated visualizations for the protection of the environment. 
}

\CCScatlist{
  \CCScatTwelve{Human-centered computing}{Visu\-al\-iza\-tion}{Visu\-al\-iza\-tion techniques}{Treemaps};
  \CCScatTwelve{Human-centered computing}{Visu\-al\-iza\-tion}{Visualization design and evaluation methods}{}
}

\begin{document}

\maketitle

\section{Introduction}
Recent surveys show that a majority of people, across countries, are increasingly worried about the deteriorating state of the environment and are ready to change their personal behavior to contribute to a better future~\cite{pew2021}. However, the consequences one can have on the environment may be unclear and not straightforward. There have been a variety of media campaigns promoting changes in behaviors which turn out to have a rather small impact, such as not sending 'thank you' emails~\cite{ovo2019emails} or unplugging one's phone charger when not in use~\cite[p.\@~ 68]{mackay2008sustainable}. Even school text books have been found not to focus on the most relevant decisions one can make in one's life~\cite{wynes2017climate}, thus leaving people confused and increasingly concerned~\cite{pew2021}. 

Our work is based on the premise that everyday-life decisions would be facilitated by a better understanding of how impact on the environment is engrained to varying degrees in (almost) everything we do and consume. We explore in this article what kind of questions related to environmental footprints one could ask while walking through an urban environment. We report preliminary results of a design exercise during which we envisioned how situated visualizations~\cite{willett2016embedded} may help people to get a better understanding of how everyday decisions and actions affect the environment and to what degree. 

\section{Background}
Approaches to support or encourage sustainable behaviors have a long tradition in the field of \emph{Environmental Psychology} \cite{vanTrijp:2013:Encouraging}, and more recently in the form of eco-feedback technology~\cite{froehlich_design_2010}  in the context of \emph{Sustainable HCI}~\cite{mankoff_environmental_2007,huang_defining_2009}. Early work focused on mechanisms to show people their resource consumption -- with the aim of thereby encouraging a reduction in consumption -- but there has been much discussion over the past years whether a focus on individual actions is justified or whether research should rather focus on changes on a societal level through systemic action, within HCI and other fields~\cite{akenji_consumer_2014,chater_i-frame_2022}. While certainly systemic actions are needed for impactful changes, we explore in this article curiosity-driven engagement with environmental questions to help people get a better sense of where impacts on the environment are caused, but also to get a better sense of quantities and how they compare. Consequently, our considerations are not directed towards persuading people to change their behavior in specific ways~\cite{brynjarsdottir_sustainably_2012}, but rather to help people surface relevant environmental data in their everyday lives.

Surfacing urban data in the form of situated visualizations has previously been explored by Vande Moere and Hill~\cite{VandeMoere:2012:UrbanData}, who discuss a range of designs, some of which are field studies exploring how people react to situated visualizations of their energy consumption, displayed publicly in the street for everyone to see~\cite{bird_pulse_2010,moere_comparative_2011}. Vande Moere and Hill also discuss a few design explorations, that is, ideas which have not been implemented as prototypes. Many imaginable designs are difficult to realize as actual prototypes due lack of accurate data or technological limitations. Here, we ignore such limitations and focus on what \emph{could} be interesting or relevant to know or learn in an urban context. Our observations may serve as directions for lifting the currently existing barriers.

\section{Situated Design Exercise}
In June 2022, a group of seven researchers set out on an in-situ ideation exercise in the French city of Bordeaux. We had previously successfully used the methodology in the context of smartwatch visualizations for sightseeing \cite{carpendale:2021:Mobile-Visualization-Design-Ideation}. Here, we adapted the methodology to the elicitation of situated visualizations in an urban environment. During a three-hour walk in the city, we strolled past several sights, busy but also lesser known streets, a park, and a restaurant. During the walk, all participants were instructed to envision information needs related to their own or others' environmental footprints and situated visualizations that would go together with these information needs. Instead of sketching out each visualization, participants took photos of the location that would hold their envisioned visualization and took notes on what it would show. In addition to serendipitous ideas, every 15--30 minutes, an alarm rang, and the whole group had to stop to devise a visualization in the current location. In contrast to the original methodology, we did not discuss ideas while in-situ but left discussions for the afternoon. We moved to a lab space in the afternoon where we collected and printed all designs and spent 1--2 hours sketching the visualizations on the photographs. These visualizations were then individually discussed and first grouped by the seven researchers involved in the activity. An independent coder who had not taken part in the activity later re-categorized the designs using the 5W1H method, that is, asking for each design the 6 questions \textit{who}, \textit{what}, \textit{why}, \textit{where}, \textit{when}, and \textit{how}.

\section{Envisioned Designs}
Overall we collected 67 different designs (corresponding to an average of approximately 10 per participant) that we attempted to group and describe more broadly. Our first categorizations centered around data and data referents\footnote{A \textit{data referent} or \textit{physical referent} is the physical object, space, or person to which the data refers \cite{willett2016embedded}. For carbon footprint data, for example, possible referents are the object or process that causes or embodies the emissions, or the person who produces, owns, or disposes of the object.} such as cars, purchases, water, or energy. The categorization in either data or referents was not entirely successful: Data and referents are closely related in some cases, but much more abstract in others. For example, a broad category was named ``places'' and included visualizations at a street corner or a park that had very different data associated with them. The street corner visualization involved showing paths to closest garbage cans or public transportation stops, while the visualization in the park showed data on plants and their history. Data and referent roughly correspond to the \textit{who}, \textit{what}, and \textit{where} questions. While those are certainly important to consider when designing a visualization, the answers to these questions did not result in an insightful classification. We therefore do not discuss them here further. Instead, we focus here on a categorization around the purposes of the situated visualizations (\textit{why}) and the mechanisms with which the situated visualizations tried to achieve their goals of aiding the environment (\textit{how}).

\subsection{Situated Visualization Purposes}

\begin{figure}[h]
    \centering
    \includegraphics[width=\columnwidth]{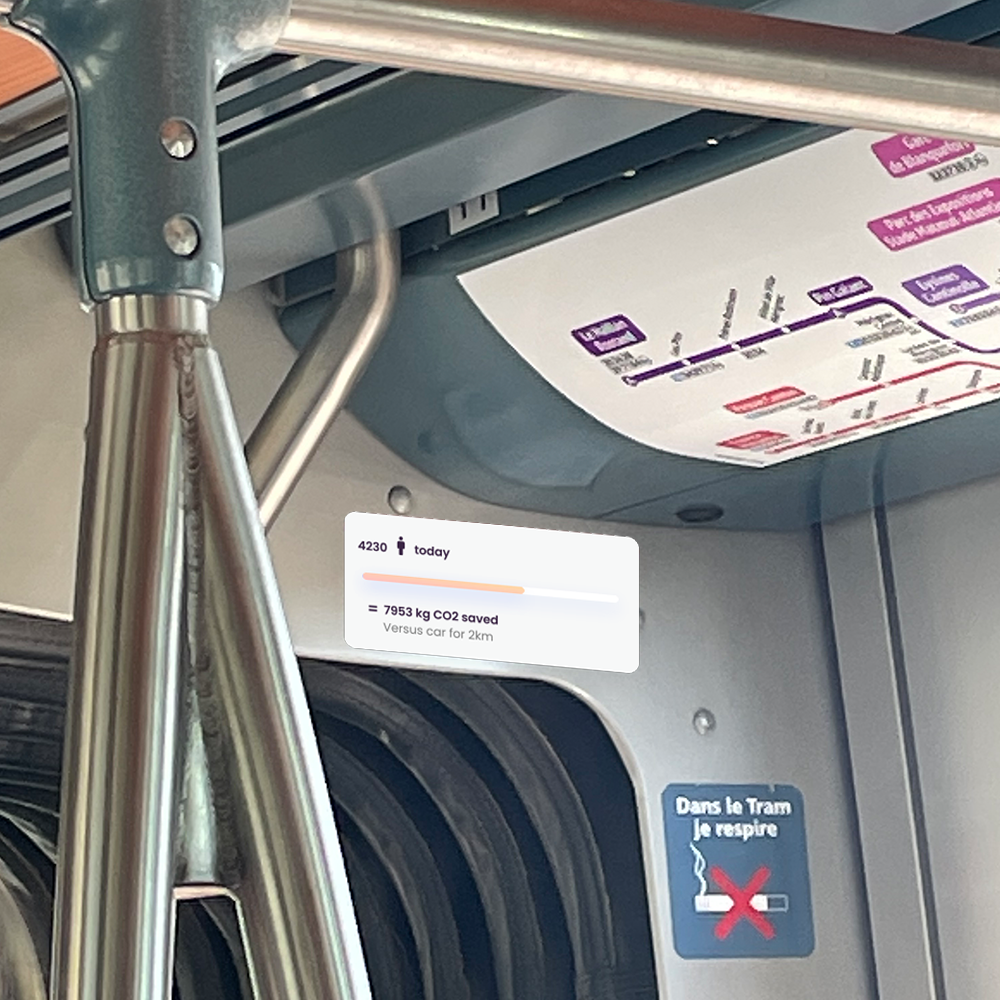}
    \caption{Example of an envisioned situated visualization facilitating eco-friendly behaviors by showing the numbers of passengers of a tramway with the amount of carbon dioxide saved compared to cars. It is placed directly inside the corresponding tramway.}
    \label{fig:ecofriendly-behavior}
\end{figure}

\textbf{Facilitating eco-friendly behavior:} Often it is hard for people to know if a specific behavior can make a non-trivial positive or negative difference on the environment, which of several behaviors would be the most eco-friendly, or how to concretely implement an eco-friendly behavior. Our corpus of ideas includes several designs meant to address these issues, such as situated visualizations showing information to help locate infrastructure, such as trash cans or public transportation stops, or to help compare the energy efficiency of public transport with individual transport options (see~\autoref{fig:ecofriendly-behavior}).

\begin{figure}[t!]
    \centering
    \includegraphics[width=\columnwidth]{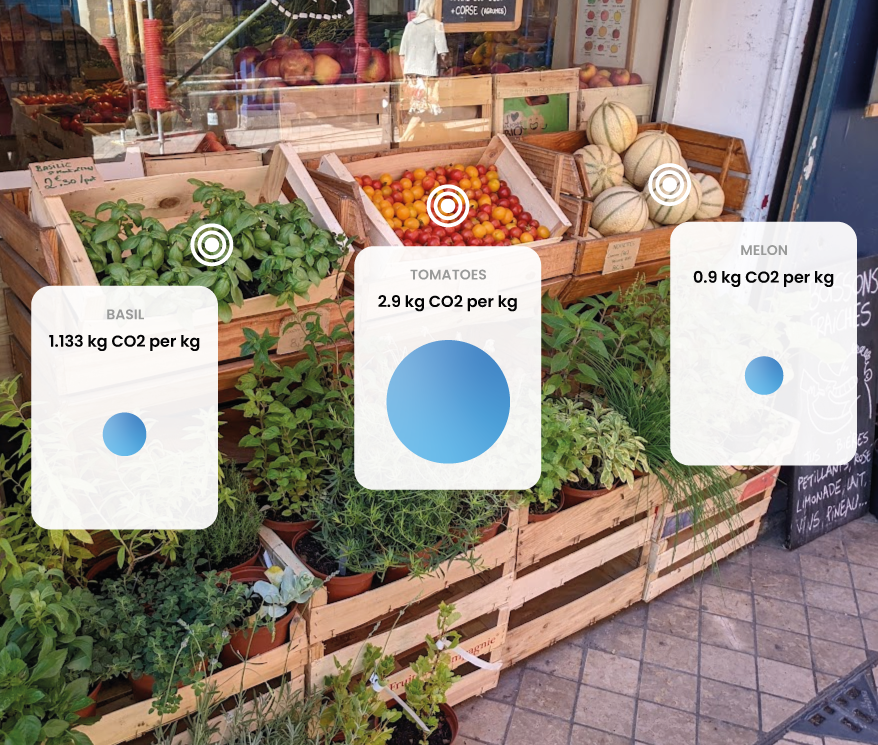}
    \caption{Example of an envisioned situated visualization informing purchasing decisions by allowing the comparison of different food items based on the carbon dioxide footprint. It is displayed directly near the corresponding food products.}
    \label{fig:purchasing-decisions}
\end{figure}

\textbf{Informing purchasing decisions:} Such visualizations are a special case of visualizations for facilitating eco-friendly behavior, but they are prominent enough to merit their own category. They consist of situated visualizations which we imagined when in a variety of situations such as next to store windows, when seeing billboard ads in the street, or when looking at restaurant menus. Concrete examples include showing visualizations of the CO$_2$ footprint of products visible in store windows (see~\autoref{fig:purchasing-decisions}) or on billboards, or the likelihood that food sold in stores or restaurants will be wasted if not bought rapidly.

\begin{figure}[b!]
    \centering
    \includegraphics[width=\columnwidth]{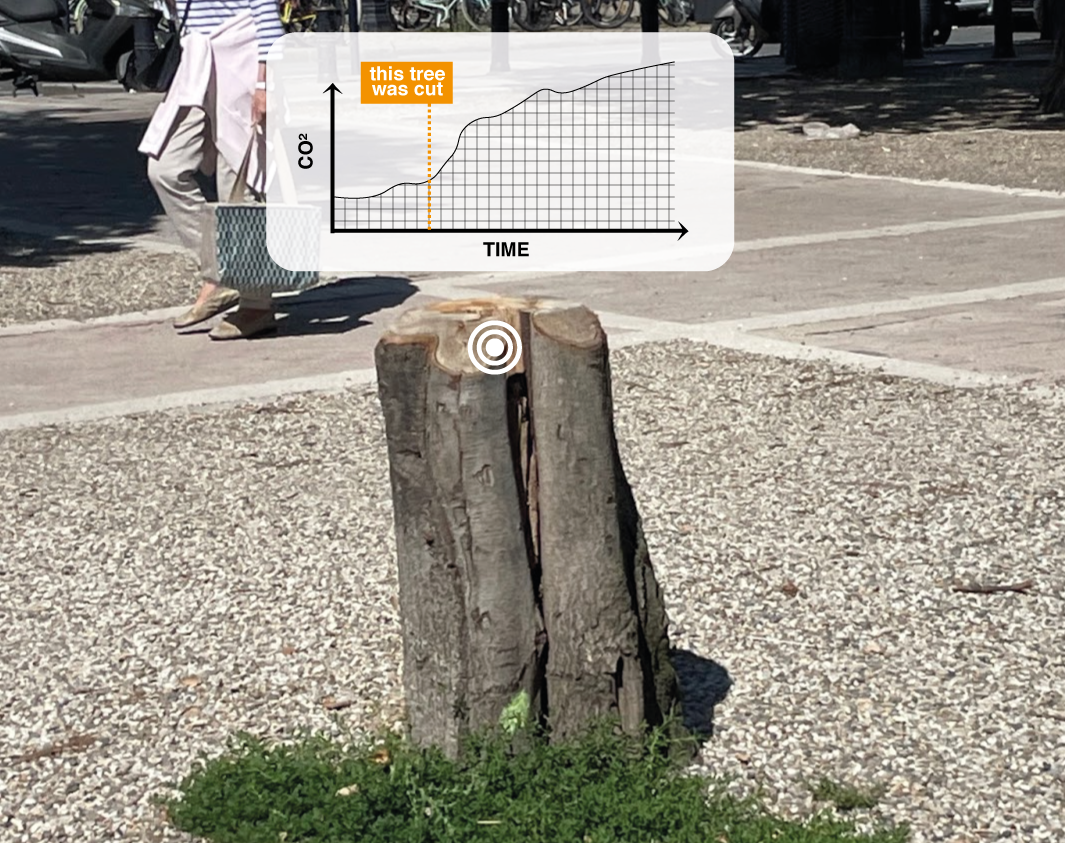}
    \caption{Example of an envisioned situated visualization increasing general environmental awareness and representing carbon dioxide produced over time. It is placed on a cut tree to emphasize trees' usefulness in reducing air pollution and increase breathability.}
    \label{fig:environmental-awareness}
\end{figure}

\textbf{Increasing general environmental awareness:} In contrast to visualizations for facilitating eco-friendly behavior and for informing purchase decisions, visualizations in this category do not aim to support immediate decision-making by viewers. Instead, they are meant to help viewers become better educated about environmental data, and get a better sense of the orders of magnitude involved. Although the knowledge viewers gain may help them make decisions later in life (e.g., vote for a specific public policy), the visualizations were not meant for immediate personal action. Examples include visualizations showing how much CO$_2$ is stored in a specific tree or how much it can capture over a year (see~\autoref{fig:environmental-awareness}).

\begin{figure}[t!]
    \centering
    \includegraphics[width=\columnwidth]{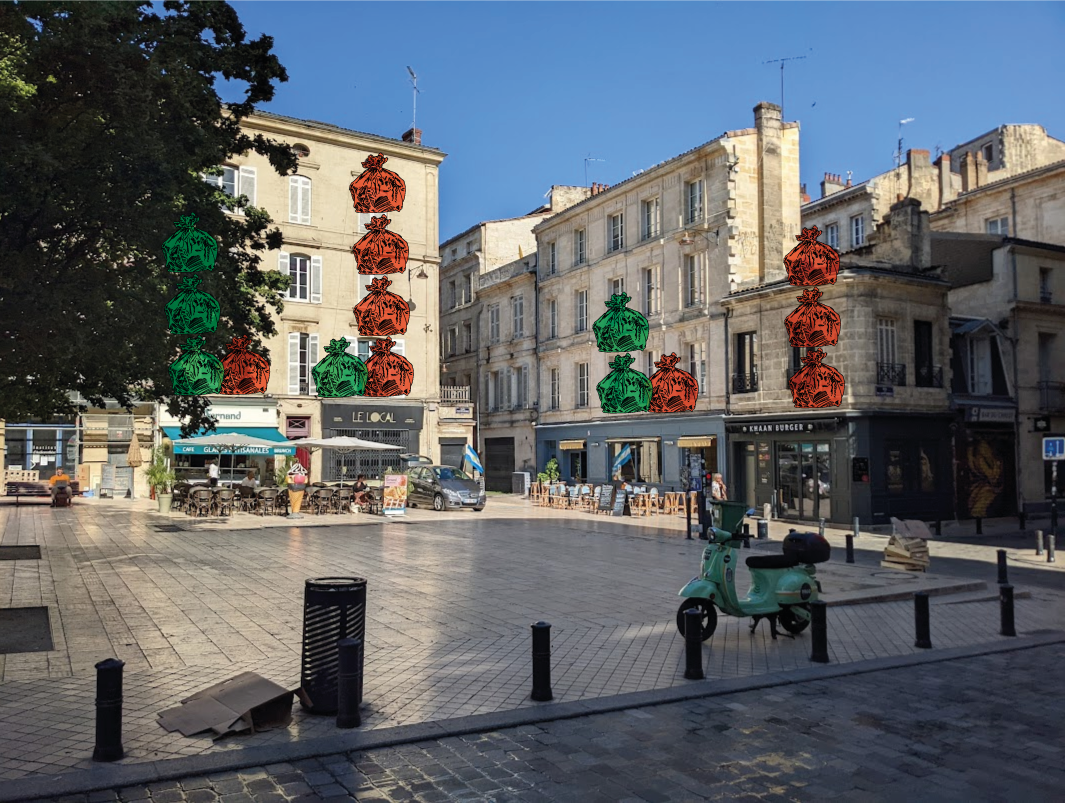}
    \caption{Example of an envisioned situated visualization increasing awareness and representing the amount of recycling trash (green) compared to regular trash (red). It is placed onto buildings from which the data is retrieved.}
    \label{fig:social-awareness}
\end{figure}

\textbf{Increasing social awareness:} These visualizations are meant to educate and increase awareness as in the previous category, but with a focus on people. They show environmental footprint data specific to individuals or groups of individuals, most often indirectly, by using objects people own or use as data referents. Examples include how much garbage was produced in each building (see~\autoref{fig:social-awareness}), as well as carbon and pollution emitted by cars, buses, and trucks, be they idling or moving on a busy road. Visualizations in this category have the ability to promote positive social feelings and behavior, but, because they can facilitate comparison between people and groups, they also have the ability to increase social pressure. 

We see as a positive aspect that some designs could fit multiple purposes. For example, one visualization shows wasted energy from a delivery van that was left running in the street while a delivery was being made. This visualization can raise awareness to passers-by about the fact that idling wastes resources but can also facilitate eco-friendly behavior when the driver sees how much unnecessary pollution was added to the air by leaving the van running. Another example, which fits the category ``informing purchasing decisions'' and ``increasing social awareness'', is a visualization in the spirit of environmental activism. Through a hidden AR visualization overlaid on advertising billboards\footnote{similar to the concept of \textit{locative art} in William Gibson's 2007 novel \textit{Spook Country}~\cite{gibson2007spook}}, it shows social data about the frequency with which other people typically replace their cell phones and why. This visualization can help people decide whether to wait to delay a phone purchase, or which models to avoid due to their lack of longevity, but can also simply educate people about the environmental impact of cell phone purchases. 

\subsection{Visualization Mechanisms}
In addition to visualization purposes, we identified a set of mechanisms with which the visualizations aimed to achieve their goals. Most situated visualizations took the form of overlays on data referents; either on infrastructure such as buildings, sculptures, or fountains or on moving objects such as cars, trams, bikes, or people. Shadows or reflections of people and objects were also used as embedding locations that were in close connection to the data referents. Other visualizations took the form of traces of referents showing historic data, for example, in the form of engravings. In addition to choices of specific embedding locations, the designs also contained a very interesting set of four mechanisms that relate to which data the visualizations show: 

\textbf{Data comparison}: Comparison was used mainly in two forms: (a) to help people compare different options, places, or objects to others and (b) to compare current data about a data referent to historic or related data. \autoref{fig:purchasing-decisions}, for example, was designed as a small multiple visualization that helps people compare food items and make purchasing decisions. Such a mechanism could help inform people concerning questions such as if buying local is better\footnote{Which is not necessarily the case if local growing of produce requires the use of greenhouses with artificial lighting}. Another visualization compares current to historic local pollution data in a street which could be a useful mechanism to evaluate reductions (or the lack of it) achieved after a change of policies, such as banning highly polluting cars. 
As such, showing comparisons was common across different purposes, be it to inform, to help make decisions, or to promote eco-friendly behavior. 

\textbf{Surface information which is by default often hidden}: Hidden information refers to data that cannot easily be inferred or extracted while in-situ. For example, how much water was needed to grow certain produce available in a store is data that is typically unavailable when making purchasing decisions. However, importantly for this category, the fact that water consumption is an issue might not even enter people's minds when buying produce. As such, surfacing information that is hidden could be a strong awareness mechanism in addition to being a purchasing aid. 

\textbf{Enhancing footprints and making them concrete:} In contrast to the previous category, some visualizations focus on enhancing data that could be experienced or inferred to a limited degree while in-situ. The two visualizations in \autoref{fig:teaser}, for example broadly concern air quality. A passer-by can, to a limited extent, sense the pollution or clarity of the air around them, but both visualizations make their respective data visually explicit and concrete. Another set of visualizations we collected was focused on the energy consumption of lights. Here, as a passer-by one sees immediately that energy is currently used to light one's path, but the amount of energy used can vary considerably depending on what type of lighting is used, and embedded visualizations could make this explicit. 

\textbf{Externalize personal knowledge for others:}  Some visualizations focused on showing data left explicitly by others in the environment. For example, situated visualizations in a dollar store represented people's experience with the longevity of the offered products. These visualizations, for example, were envisioned in the form of hidden AR graffiti or annotations of products or places and were meant to help decision making\footnote{We imagined them hidden similar to the billboard overlay mentioned before since store owners may object to the externalization of such information. See section 5 for a short discussion of this potential issue.}.

\section{Future Research and Conclusions}
In this ideation exercise we have done a first step towards envisioning visualizations of various impacts of humans on the urban environment. For the very short timeframe of the exercise, essentially one work day, we elicited a rich set of ideas that can each be explored further. Some visualizations aimed to aid people identify which actions are in line with their stated attitudes~\cite{pew2021}, for example, by helping make purchasing decisions or finding their way to the next local transport stop. Others were of purely informational nature, intended to help people get a sense of orders of magnitude for carbon emissions and sequestration, such as showing the amount of carbon sequestered in a tree. Other visualizations were trying to convince people, for example, by showing environmentally conscious decisions of others such as commuting by bike instead of by car or delaying phone purchases. Some visualizations went even in the direction of environmental activism in that they overlaid information on company logos or advertisements to raise awareness of the negative impact of specific companies on the environment. Like these, some of the envisioned designs that would enrich products with eco-footprint visualizations would likely clash with business interests of companies. One could argue though that in the end, surfacing such data may actually be in the interest of the company: the market is driven by demand; reducing the demand for polluting products and increasing demand for less polluting ones might help a company to shift their production in the right direction.

Much remains to be explored to study how, why, where, and when situated visualizations can best help the environment. Many of the visualizations conceived during the described exercise would currently be difficult to deploy due to issues such as technological barriers or data availability. Some of the examples would also require the installation of new infrastructure, which risks introducing additional footprints. In such cases, one has to ask if that is indeed necessary or if a similar effect could be achieved using existing technology, such as smartphones which are already ubiquitous. For fixed installations, one would also need to consider ``human'' issues such as theft or vandalism. 
Finally, we also need to think about the cognitive impact on viewers when many visualisations appear in their field of view. Information overload may be a real problem especially when viewers are already engaged in an activity that is cognitively intense. Yet, these words of caution should not distract from starting to work on situated visualizations to help the environment. 

Determining whether situated visualizations of ecological footprints are actually useful can be difficult, though. While one should make sure that the visualization design used can be correctly interpreted by viewers -- as we do whenever we evaluate how long it takes people to extract information correctly from a given visualization -- we will need a different type of measure to determine if our situated visualizations can be considered effective. One possibility could be long-term behavior change studies, but one has to ask, how long is enough? While climate change is a slow process, we reached a point where we need to act quickly to stay within favorable scenarios~\cite{ippc6}. It would thus be preferable to find measures that enabled us to determine the effectiveness of an approach without spending a year on each iteration. Approaches taken in other domains could be an option and their applicability needs to be explored. For example, economists argue that the most important tool to move our societies to carbon-neutrality is the introduction of a carbon tax priced such that emission goals are met\footnote{As detailed in this statement written up by well-known economists and signed by many more since \url{https://www.econstatement.org/}.}. A 2019 article built on this and explored across multiple pre-registered experiments how comparatively less effective ``Green Nudges'' (such as defaulting people to a green energy provider) reduced participants' support for such a carbon tax~\cite{hagmann_nudging_2019}. That same study also suggests that providing people with better information about the tax reduced the negative impact of the availability of the nudge. We conclude by highlighting that there is a rich area to be explored at the intersection of visualization, economics, and psychology to find new ways to help people engage with and understand the complex and often unintuitive relationships between natural and human-made phenomena affecting the environment, and to determine if these new ways can have an impact. 

\balance
\acknowledgments{
This work was partly supported by the Agence Nationale de la Recherche (ANR), grant number ANR-19-CE33-0012.}

\bibliographystyle{abbrv-doi-hyperref}

\bibliography{emberworkshop}
\end{document}


%% file: Ember.bbl
\begin{thebibliography}{10}

\bibitem{akenji_consumer_2014}
\href{https://doi.org/10.1016/j.jclepro.2013.05.022}{L.~Akenji}.
\newblock \href{https://doi.org/10.1016/j.jclepro.2013.05.022}{Consumer
  scapegoatism and limits to green consumerism}.
\newblock \href{https://doi.org/10.1016/j.jclepro.2013.05.022}{{\em Journal of
  Cleaner Production}},
  \href{https://doi.org/10.1016/j.jclepro.2013.05.022}{63:13--23},
  \href{https://doi.org/10.1016/j.jclepro.2013.05.022}{Jan. 2014}.
  \href{https://doi.org/10.1016/j.jclepro.2013.05.022}
{doi: {{%
10\hspace{.1pt}\discretionary{.}{%
}{.}\hspace{.4pt}1016\discretionary{/}{%
}{/}j\hspace{.1pt}\discretionary{.}{%
}{.}\hspace{.4pt}jclepro\hspace{.1pt}\discretionary{.}{%
}{.}\hspace{.4pt}2013\hspace{.1pt}\discretionary{.}{%
}{.}\hspace{.4pt}05\hspace{.1pt}\discretionary{.}{%
}{.}\hspace{.4pt}022}}}


\bibitem{pew2021}
J.~Bell, J.~Poushter, M.~Fagan, and C.~Huang.
\newblock In response to climate change, citizens in advanced economies are
  willing to alter how they live and work.
\newblock Online report, September 2021.
\newblock
  \href{https://www.pewresearch.org/global/2021/09/14/in-response-to-climate-change-citizens-in-advanced-economies-are-willing-to-alter-how-they-live-and-work/}{https://www.pewresearch.org/global/2021/09/14/in-response-to-climate-change-citizens-in-advanced-economies-are-willing-to-alter-how-they-live-and-work/}.

\bibitem{bird_pulse_2010}
J.~Bird and Y.~Rogers.
\newblock The pulse of tidy street: {Measuring} and publicly displaying
  domestic electricity consumption.
\newblock In {\em Workshop on energy awareness and conservation through
  pervasive applications (held at {Pervasive})}, 2010.

\bibitem{brynjarsdottir_sustainably_2012}
\href{https://doi.org/10.1145/2207676.2208539}{H.~Brynjarsdóttir,
  M.~Håkansson, J.~Pierce, E.~P.~S. Baumer, C.~DiSalvo, and P.~Sengers}.
\newblock \href{https://doi.org/10.1145/2207676.2208539}{Sustainably
  unpersuaded: how persuasion narrows our vision of sustainability}.
\newblock \href{https://doi.org/10.1145/2207676.2208539}{In {\em Proceedings of
  the {Conference} on {Human} {Factors} in {Computing} {Systems} (CHI)}},
  \href{https://doi.org/10.1145/2207676.2208539}{pp. 947--956},
  \href{https://doi.org/10.1145/2207676.2208539}{2012}.
  \href{https://doi.org/10.1145/2207676.2208539}
{doi: {{%
10\hspace{.1pt}\discretionary{.}{%
}{.}\hspace{.4pt}1145\discretionary{/}{%
}{/}2207676\hspace{.1pt}\discretionary{.}{%
}{.}\hspace{.4pt}2208539}}}


\bibitem{carpendale:2021:Mobile-Visualization-Design-Ideation}
\href{https://doi.org/https://doi.org/10.1201/9781003090823-8}{S.~Carpendale,
  P.~Isenberg, C.~Perin, T.~Blascheck, F.~Daneshzand, A.~Islam, K.~Currier,
  P.~Buk, V.~Cheung, L.~Quach, and L.~Vermette}.
\newblock \href{https://doi.org/https://doi.org/10.1201/9781003090823-8}{Mobile
  visualization design: An ideation method to try}.
\newblock \href{https://doi.org/https://doi.org/10.1201/9781003090823-8}{In
  {\em {Mobile Data Visualization}}},
  \href{https://doi.org/https://doi.org/10.1201/9781003090823-8}{pp. 241--261}.
  \href{https://doi.org/https://doi.org/10.1201/9781003090823-8}{{Chapman and
  Hall/CRC}},
  \href{https://doi.org/https://doi.org/10.1201/9781003090823-8}{Nov. 2021}.
  \href{https://doi.org/10.1201/9781003090823-8}
{doi: {{%
10\hspace{.1pt}\discretionary{.}{%
}{.}\hspace{.4pt}1201\discretionary{/}{%
}{/}9781003090823\discretionary{%
}{-}{-}8}}}


\bibitem{chater_i-frame_2022}
\href{https://doi.org/10.2139/ssrn.4046264}{N.~Chater and G.~Loewenstein}.
\newblock \href{https://doi.org/10.2139/ssrn.4046264}{The i-{Frame} and the
  s-{Frame}: {How} focusing on individual-level solutions has led behavioral
  public policy astray}.
\newblock \href{https://doi.org/10.2139/ssrn.4046264}{{SSRN} {Scholarly}
  {Paper} 4046264}, \href{https://doi.org/10.2139/ssrn.4046264}{Mar. 2022}.
  \href{https://doi.org/10.2139/ssrn.4046264}
{doi: {{%
10\hspace{.1pt}\discretionary{.}{%
}{.}\hspace{.4pt}2139\discretionary{/}{%
}{/}ssrn\hspace{.1pt}\discretionary{.}{%
}{.}\hspace{.4pt}4046264}}}


\bibitem{froehlich_design_2010}
\href{https://doi.org/10.1145/1753326.1753629}{J.~Froehlich, L.~Findlater, and
  J.~Landay}.
\newblock \href{https://doi.org/10.1145/1753326.1753629}{The design of
  eco-feedback technology}.
\newblock \href{https://doi.org/10.1145/1753326.1753629}{In {\em Proceedings of
  the {Conference} on {Human} {Factors} in {Computing} {Systems} (CHI)}},
  \href{https://doi.org/10.1145/1753326.1753629}{pp. 1999--2008}.
  \href{https://doi.org/10.1145/1753326.1753629}{New York, NY, USA},
  \href{https://doi.org/10.1145/1753326.1753629}{2010}.
  \href{https://doi.org/10.1145/1753326.1753629}
{doi: {{%
10\hspace{.1pt}\discretionary{.}{%
}{.}\hspace{.4pt}1145\discretionary{/}{%
}{/}1753326\hspace{.1pt}\discretionary{.}{%
}{.}\hspace{.4pt}1753629}}}


\bibitem{gibson2007spook}
W.~Gibson.
\newblock {\em Spook country}.
\newblock Penguin, 2007.

\bibitem{hagmann_nudging_2019}
\href{https://doi.org/10.1038/s41558-019-0474-0}{D.~Hagmann, E.~H. Ho, and
  G.~Loewenstein}.
\newblock \href{https://doi.org/10.1038/s41558-019-0474-0}{Nudging out support
  for a carbon tax}.
\newblock \href{https://doi.org/10.1038/s41558-019-0474-0}{{\em Nature Climate
  Change}}, \href{https://doi.org/10.1038/s41558-019-0474-0}{9(6):484--489},
  \href{https://doi.org/10.1038/s41558-019-0474-0}{2019}.
\newblock \href{https://doi.org/10.1038/s41558-019-0474-0}{Publisher: Nature
  Publishing Group}. \href{https://doi.org/10.1038/s41558-019-0474-0}
{doi: {{%
10\hspace{.1pt}\discretionary{.}{%
}{.}\hspace{.4pt}1038\discretionary{/}{%
}{/}s41558\discretionary{%
}{-}{-}019\discretionary{%
}{-}{-}0474\discretionary{%
}{-}{-}0}}}


\bibitem{huang_defining_2009}
\href{https://doi.org/10.1145/1520340.1520751}{E.~M. Huang, E.~Blevis,
  J.~Mankoff, L.~P. Nathan, and B.~Tomlinson}.
\newblock \href{https://doi.org/10.1145/1520340.1520751}{Defining the role of
  {HCI} in the challenges of sustainability}.
\newblock \href{https://doi.org/10.1145/1520340.1520751}{In {\em Extended
  Abstracts of the Conference on Human Factors in Computing Systems (CHI)}},
  \href{https://doi.org/10.1145/1520340.1520751}{pp. 4827--4830}.
  \href{https://doi.org/10.1145/1520340.1520751}{2009}.
  \href{https://doi.org/10.1145/1520340.1520751}
{doi: {{%
10\hspace{.1pt}\discretionary{.}{%
}{.}\hspace{.4pt}1145\discretionary{/}{%
}{/}1520340\hspace{.1pt}\discretionary{.}{%
}{.}\hspace{.4pt}1520751}}}


\bibitem{mackay2008sustainable}
\href{https://www.withouthotair.com/c11/page_68.shtml}{D.~MacKay}.
\newblock \href{https://www.withouthotair.com/c11/page_68.shtml}{{\em
  Sustainable Energy --- Without the Hot Air}}.
\newblock \href{https://www.withouthotair.com/c11/page_68.shtml}{UIT
  cambridge}, \href{https://www.withouthotair.com/c11/page_68.shtml}{2008}.

\bibitem{mankoff_environmental_2007}
\href{https://doi.org/10.1145/1240866.1240963}{J.~C. Mankoff, E.~Blevis,
  A.~Borning, B.~Friedman, S.~R. Fussell, J.~Hasbrouck, A.~Woodruff, and
  P.~Sengers}.
\newblock \href{https://doi.org/10.1145/1240866.1240963}{Environmental
  sustainability and interaction}.
\newblock \href{https://doi.org/10.1145/1240866.1240963}{In {\em Extended
  Abstracts of the Conference on Human Factors in Computing Systems (CHI)}},
  \href{https://doi.org/10.1145/1240866.1240963}{pp. 2121--2124},
  \href{https://doi.org/10.1145/1240866.1240963}{2007}.
  \href{https://doi.org/10.1145/1240866.1240963}
{doi: {{%
10\hspace{.1pt}\discretionary{.}{%
}{.}\hspace{.4pt}1145\discretionary{/}{%
}{/}1240866\hspace{.1pt}\discretionary{.}{%
}{.}\hspace{.4pt}1240963}}}


\bibitem{ovo2019emails}
\href{https://www.ovoenergy.com/ovo-newsroom/press-releases/2019/november/think-before-you-thank-if-every-brit-sent-one-less-thank-you-email-a-day-we-would-save-16433-tonnes-of-carbon-a-year-the-same-as-81152-flights-to-madrid}{{OVO
  Energy}}.
\newblock
  \href{https://www.ovoenergy.com/ovo-newsroom/press-releases/2019/november/think-before-you-thank-if-every-brit-sent-one-less-thank-you-email-a-day-we-would-save-16433-tonnes-of-carbon-a-year-the-same-as-81152-flights-to-madrid}{‘think
  before you thank’: If every brit sent one less thank you email a day, we
  would save 16,433 tonnes of carbon a year - the same as 81,152 flights to
  madrid}.
\newblock
  \href{https://www.ovoenergy.com/ovo-newsroom/press-releases/2019/november/think-before-you-thank-if-every-brit-sent-one-less-thank-you-email-a-day-we-would-save-16433-tonnes-of-carbon-a-year-the-same-as-81152-flights-to-madrid}{Online
  press release},
  \href{https://www.ovoenergy.com/ovo-newsroom/press-releases/2019/november/think-before-you-thank-if-every-brit-sent-one-less-thank-you-email-a-day-we-would-save-16433-tonnes-of-carbon-a-year-the-same-as-81152-flights-to-madrid}{2019}.
\newblock
  \href{https://www.ovoenergy.com/ovo-newsroom/press-releases/2019/november/think-before-you-thank-if-every-brit-sent-one-less-thank-you-email-a-day-we-would-save-16433-tonnes-of-carbon-a-year-the-same-as-81152-flights-to-madrid}{https://www.ovoenergy.com/ovo-newsroom/press-releases/2019/november/think-before-you-thank-if-every-brit-sent-one-less-thank-you-email-a-day-we-would-save-16433-tonnes-of-carbon-a-year-the-same-as-81152-flights-to-madrid}.

\bibitem{ippc6}
\href{https://doi.org/10.1017/9781009157926}{P.~Shukla, J.~Skea, R.~Slade,
  A.~A. Khourdajie, R.~van Diemen, D.~McCollum, M.~Pathak, S.~Some, P.~Vyas,
  R.~Fradera, M.~Belkacemi, A.~Hasija, G.~Lisboa, S.~Luz, and
  J.~Malley~[eds.]}.
\newblock \href{https://doi.org/10.1017/9781009157926}{{\em IPCC, 2022: Climate
  Change 2022: Mitigation of Climate Change. Contribution of Working Group III
  to the Sixth Assessment Report of the Intergovernmental Panel on Climate
  Change}}.
\newblock \href{https://doi.org/10.1017/9781009157926}{Cambridge University
  Press, Cambridge, UK and New York, NY, USA},
  \href{https://doi.org/10.1017/9781009157926}{2022}.
  \href{https://doi.org/10.1017/9781009157926}
{doi: {{%
10\hspace{.1pt}\discretionary{.}{%
}{.}\hspace{.4pt}1017\discretionary{/}{%
}{/}9781009157926}}}


\bibitem{vanTrijp:2013:Encouraging}
\href{https://doi.org/10.4324/9780203141182}{H.~C. van Trijp, ed.}
\newblock \href{https://doi.org/10.4324/9780203141182}{{\em Encouraging
  Sustainable Behavior: Psychology and the Environment}}.
\newblock \href{https://doi.org/10.4324/9780203141182}{Psychology Press},
  \href{https://doi.org/10.4324/9780203141182}{2013}.
  \href{https://doi.org/10.4324/9780203141182}
{doi: {{%
10\hspace{.1pt}\discretionary{.}{%
}{.}\hspace{.4pt}4324\discretionary{/}{%
}{/}9780203141182}}}


\bibitem{VandeMoere:2012:UrbanData}
\href{https://doi.org/10.1080/10630732.2012.698065}{A.~Vande~Moere and
  D.~Hill}.
\newblock \href{https://doi.org/10.1080/10630732.2012.698065}{Designing for the
  situated and public visualization of urban data}.
\newblock \href{https://doi.org/10.1080/10630732.2012.698065}{{\em Journal of
  Urban Technology}},
  \href{https://doi.org/10.1080/10630732.2012.698065}{19(2):25--46},
  \href{https://doi.org/10.1080/10630732.2012.698065}{2012}.
  \href{https://doi.org/10.1080/10630732.2012.698065}
{doi: {{%
10\hspace{.1pt}\discretionary{.}{%
}{.}\hspace{.4pt}1080\discretionary{/}{%
}{/}10630732\hspace{.1pt}\discretionary{.}{%
}{.}\hspace{.4pt}2012\hspace{.1pt}\discretionary{.}{%
}{.}\hspace{.4pt}698065}}}


\bibitem{moere_comparative_2011}
\href{https://doi.org/10.1007/978-3-642-23774-4_39}{A.~Vande~Moere,
  M.~Tomitsch, M.~Hoinkis, E.~Trefz, S.~Johansen, and A.~Jones}.
\newblock \href{https://doi.org/10.1007/978-3-642-23774-4_39}{Comparative
  feedback in the street: exposing residential energy consumption on house
  façades}.
\newblock \href{https://doi.org/10.1007/978-3-642-23774-4_39}{In {\em {IFIP}
  {Conference} on {Human}-{Computer} {Interaction} (INTERACT)}},
  \href{https://doi.org/10.1007/978-3-642-23774-4_39}{pp. 470--488},
  \href{https://doi.org/10.1007/978-3-642-23774-4_39}{2011}.
  \href{https://doi.org/10.1007/978-3-642-23774-4_39}
{doi: {{%
10\hspace{.1pt}\discretionary{.}{%
}{.}\hspace{.4pt}1007\discretionary{/}{%
}{/}978\discretionary{%
}{-}{-}3\discretionary{%
}{-}{-}642\discretionary{%
}{-}{-}23774\discretionary{%
}{-}{-}4\_39}}}


\bibitem{willett2016embedded}
\href{https://doi.org/10.1109/TVCG.2016.2598608}{W.~Willett, Y.~Jansen, and
  P.~Dragicevic}.
\newblock \href{https://doi.org/10.1109/TVCG.2016.2598608}{Embedded data
  representations}.
\newblock \href{https://doi.org/10.1109/TVCG.2016.2598608}{{\em {IEEE
  Transactions on Visualization and Computer Graphics}}},
  \href{https://doi.org/10.1109/TVCG.2016.2598608}{23(1):461--470},
  \href{https://doi.org/10.1109/TVCG.2016.2598608}{2016}.
  \href{https://doi.org/10.1109/TVCG.2016.2598608}
{doi: {{%
10\hspace{.1pt}\discretionary{.}{%
}{.}\hspace{.4pt}1109\discretionary{/}{%
}{/}TVCG\hspace{.1pt}\discretionary{.}{%
}{.}\hspace{.4pt}2016\hspace{.1pt}\discretionary{.}{%
}{.}\hspace{.4pt}2598608}}}


\bibitem{wynes2017climate}
\href{https://doi.org/10.1088/1748-9326/aa7541}{S.~Wynes and K.~A. Nicholas}.
\newblock \href{https://doi.org/10.1088/1748-9326/aa7541}{The climate
  mitigation gap: education and government recommendations miss the most
  effective individual actions}.
\newblock \href{https://doi.org/10.1088/1748-9326/aa7541}{{\em Environmental
  Research Letters}},
  \href{https://doi.org/10.1088/1748-9326/aa7541}{12(7):074024},
  \href{https://doi.org/10.1088/1748-9326/aa7541}{2017}.
  \href{https://doi.org/10.1088/1748-9326/aa7541}
{doi: {{%
10\hspace{.1pt}\discretionary{.}{%
}{.}\hspace{.4pt}1088\discretionary{/}{%
}{/}1748\discretionary{%
}{-}{-}9326\discretionary{/}{%
}{/}aa7541}}}


\end{thebibliography}
